\documentstyle[12pt]{article}

\catcode`\@=11
\long\def\@makefntext#1{
\protect\noindent \hbox to 3.2pt {\hskip-.9pt
$^{{\ninerm\@thefnmark}}$\hfil}#1\hfill}                %CAN BE USED

 \def\@makefnmark{\hbox to 0pt{$^{\@thefnmark}$\hss}}  %ORIGINAL

\def\ps@myheadings{\let\@mkboth\@gobbletwo
\def\@oddhead{\hbox{}
\rightmark\hfil\ninerm\thepage}
\def\@oddfoot{}\def\@evenhead{\ninerm\thepage\hfil
\leftmark\hbox{}}\def\@evenfoot{}
\def\sectionmark##1{}\def\subsectionmark##1{}}

\newcounter{sectionc}\newcounter{subsectionc}\newcounter{subsubsectionc}
\renewcommand{\section}[1] {\vspace{0.6cm}\addtocounter{sectionc}{1}
\setcounter{subsectionc}{0}\setcounter{subsubsectionc}{0}\noindent
	{\bf\thesectionc. #1}\par\vspace{0.4cm}}
\renewcommand{\subsection}[1] {\vspace{0.6cm}\addtocounter{subsectionc}{1}
	\setcounter{subsubsectionc}{0}\noindent
	{\it\thesectionc.\thesubsectionc. #1}\par\vspace{0.4cm}}
\renewcommand{\subsubsection}[1]
{\vspace{0.6cm}\addtocounter{subsubsectionc}{1}
	\noindent {\rm\thesectionc.\thesubsectionc.\thesubsubsectionc.
	#1}\par\vspace{0.4cm}}

\newcounter{appendixc}
\newcounter{subappendixc}[appendixc]
\newcounter{subsubappendixc}[subappendixc]

\renewcommand{\appendix}[1] {\vspace{0.6cm}
	\refstepcounter{appendixc}
	\setcounter{figure}{0}
	\setcounter{table}{0}
	\setcounter{equation}{0}
	\renewcommand{\thefigure}{\Alph{appendixc}.\arabic{figure}}
	\renewcommand{\thetable}{\Alph{appendixc}.\arabic{table}}
	\renewcommand{\theappendixc}{\Alph{appendixc}}
	\renewcommand{\theequation}{\Alph{appendixc}.\arabic{equation}}
	\noindent{\bf Appendix \theappendixc #1}\par\vspace{0.4cm}}

\renewenvironment{thebibliography}[1]
	{\begin{list}{\arabic{enumi}.}
	{\usecounter{enumi}\setlength{\parsep}{0pt}
\setlength{\leftmargin 1.25cm}{\rightmargin 0pt}
	 \setlength{\itemsep}{0pt} \settowidth
	{\labelwidth}{#1.}\sloppy}}{\end{list}}

\newcounter{itemlistc}
\newcounter{romanlistc}
\newcounter{alphlistc}
\newcounter{arabiclistc}

\newcommand{\fcaption}[1]{
	\refstepcounter{figure}
	\setbox\@tempboxa = \hbox{\tenrm Fig.~\thefigure. #1}
	\ifdim \wd\@tempboxa > 6in
	   {\begin{center}
	\parbox{6in}{\tenrm\baselineskip=12pt Fig.~\thefigure. #1}
	    \end{center}}
	\else
	     {\begin{center}
	     {\tenrm Fig.~\thefigure. #1}
	      \end{center}}
	\fi}

\newcommand{\tcaption}[1]{
	\refstepcounter{table}
	\setbox\@tempboxa = \hbox{\tenrm Table~\thetable. #1}
	\ifdim \wd\@tempboxa > 6in
	   {\begin{center}
	\parbox{6in}{\tenrm\baselineskip=12pt Table~\thetable. #1}
	    \end{center}}
	\else
	     {\begin{center}
	     {\tenrm Table~\thetable. #1}
	      \end{center}}
	\fi}

\def\@citex[#1]#2{\if@filesw\immediate\write\@auxout
	{\string\citation{#2}}\fi
\def\@citea{}\@cite{\@for\@citeb:=#2\do
	{\@citea\def\@citea{,}\@ifundefined
	{b@\@citeb}{{\bf ?}\@warning
	{Citation `\@citeb' on page \thepage \space undefined}}
	{\csname b@\@citeb\endcsname}}}{#1}}

\newif\if@cghi
\def\cite{\@cghitrue\@ifnextchar [{\@tempswatrue
	\@citex}{\@tempswafalse\@citex[]}}
\def\citelow{\@cghifalse\@ifnextchar [{\@tempswatrue
	\@citex}{\@tempswafalse\@citex[]}}
\def\@cite#1#2{{$\null^{#1}$\if@tempswa\typeout
	{IJCGA warning: optional citation argument
	ignored: `#2'} \fi}}

\def\fnt#1#2{\footnotetext{\kern-.3em
	{$^{\mbox{\sevenrm #1}}$}{#2}}}

 1
 1
 1

\font\tenrm=cmr10

\font\ninerm=cmr9

\newcommand{\bea}{\begin{eqnarray}}
\newcommand{\eea}{\end{eqnarray}}
\newcommand{\beq}{\begin{equation}}

\newcommand{\eeq}{\end{equation}}
\newcommand{\bec}{\begin{center}}
\newcommand{\eec}{\end{center}}
\newcommand{\beqn}{\begin{displaymath}}
\newcommand{\eeqn}{\end{displaymath}}
\newcommand{\zeilea}{\renewcommand{\baselinestretch}{1.5}
		     \small\normalsize}
\newcommand{\zeilee}{\renewcommand{\baselinestretch}{1}
		     \small\normalsize}
\newcommand{\vslash}{\mbox{$\not{\hspace{-0.7mm}v}$}}           % vslash

\newcommand{\eins}{\mbox{$\rule{2.5mm}{0.1mm}
			  {\hspace{-2.7mm}1}
			  {\hspace{-0.2mm}\rule{0.07mm}{2.7mm}}$}}
\newcommand{\mini}{\mbox{\scriptsize min}}
\newcommand{\Yk}  {\begin{Young}
		      $\scriptstyle{k}$\cr
		      \end{Young} }
\newcommand{\YK}  {\begin{Young}
		      $\scriptscriptstyle{K}$\cr
		      \end{Young} }

\newcommand{\Yzehn}{\begin{Young}
		      &\cr
		      \end{Young} }
\newcommand{\Yeins}  {\begin{Young}
		      \cr
		      \cr
		      \cr
		      \end{Young} }
\newcommand{\Yanti}    {\begin{Young}
		      \cr
		      \cr
		      \end{Young} }
\newdimen\hoogte    \hoogte=7pt    % hoogte  van hokje
\newdimen\breedte   \breedte=9pt   % breedte van hokje
\newdimen\dikte     \dikte=0.5pt    % dikte lijn
\def\beginYoung{
       \begingroup
       \def\vr{\vrule height0.8\hoogte width\dikte depth 0.2\hoogte}
       \def\fbox##1{\vbox{\offinterlineskip
		    \hrule height\dikte
		    \hbox to \breedte{\vr\hfill##1\hfill\vr}
		    \hrule height\dikte}}
       \vbox\bgroup \offinterlineskip \tabskip=-\dikte \lineskip=-\dikte
	    \halign\bgroup &\fbox{##\unskip}\unskip  \crcr }
\def\End@Young{\egroup\egroup\endgroup}
\newenvironment{Young}{\beginYoung}{\End@Young}
\begin{document}

\begin{titlepage}
\begin{flushright}
 IC/95/45\\
 MZ-TH/95-15\\
 hep-ph/9505335
\end{flushright}
\vspace*{0.5cm}
\begin{center}
{\Huge $SU(2N_{f})\otimes O(3)$ light diquark symmetry and
current-induced heavy baryon transition form factors}

\vspace*{0.5cm}
{\large F. Hussain}

{\large International Centre for Theoretical Physics, Trieste, Italy}\\
\vspace*{0.5cm}
{\large J.G. K\"orner\footnote[1]
{Supported in part by  the BMFT, FRG, under contract 06MZ730},
J. Landgraf\footnotemark[1] }\\
{\large Institut f\"ur Physik, Johannes Gutenberg-Universit\"at,\\
 Staudinger Weg 7, D-55099 Mainz, Germany}\\
\vspace{0.5cm}
{\large and}\\
\vspace*{0.5cm}
{\large Salam Tawfiq}\\
{\large Dipartimento di Fisica Teorica, Universit\`a degli Studi \\ di
Trieste, Trieste, Italy }\\
\vspace{0.5cm}
{\large May 1995}
\vspace*{0.5cm}
\begin{abstract}
We study the current-induced bottom baryon to charm baryon transitions
in the Heavy Quark Symmetry limit as $m_{q}\rightarrow \infty$. Our
discussion involves $s$-wave to $s$-wave as well as $s$-wave to $p$-wave
transitions.
Using a constituent quark model picture
for the light diquark system with an underlying $SU(2N_{f})\otimes O(3)$
symmetry and the heavy quark symmetry we arrive at a number of new
predictions for the reduced form factors that describe these transitions.
\end{abstract}
\end{center}
\end{titlepage}

\section{Introduction}
The Heavy Quark Effective Theory (HQET) formulated in 1990\cite{C1} is
so well known by now that it no longer needs an extensive introduction. The
HQET provides a systematic expansion of QCD in terms of inverse powers of
the heavy quark mass. The leading term in this expansion gives rise to a
new spin  and flavour symmetry at equal velocities, termed the Heavy Quark
Symmetry. Corrections to the Heavy Quark Symmetry limit can be classified and
evaluated order by order in $1/m_Q$ by considering the contributions of the
nonleading terms
in the effective HQET fields and the HQET Langrangian.

In this paper we will only be concerned with the leading order term in the
HQET expansion, i.e. in the Heavy Quark Symmetry limit. In this limit the
dynamics of the light and heavy constituents decouple and the calculation
of the Heavy Quark Symmetry predictions for transition amplitudes
essentially amounts to an angular momentum coupling exercise, however
involved it may be. It is then not surprising that such calculations have
been done even before the conceptual foundations of HQET had been laid
down in 1990. For example, the Heavy Quark Symmetry structure of
current-induced charm baryon transitions had been written down as early as
1976\cite{C5}. As concerns the angular momentum coupling
calculations these can either be done using Clebsch-Gordan coefficents or, more
compactly, using the Wigner 6-$j$-symbol calculus. Still a different method
is to use covariant spin wave functions or the Bethe-Salpeter (B-S) formalism.
In our previous works, Refs. [3]-[7], we have used this formalism to
derive the
consequences of the heavy quark symmetry for weak transitions of hadrons
of arbitrary spin.   This is the approach taken
by most workers in the field and as well as in this paper.

At present most of the attention of experimentalists and theoreticians
working in heavy quark physics is directed towards the application of HQET
in the meson sector where data is starting to become quite abundant. This
data will be supplemented in the not-too-distant future by corresponding
data on heavy baryon decays and there will be a need to analyze this data
in terms of HQET as applied to heavy baryons. In the present work we
disuss the structure of flavour changing bottom baryon to charm baryon
decays in the Heavy Quark Symmetry limit as $m_{Q}\rightarrow \infty$.
We go beyond our previous works on the subject in that we
treat the light diquark system in a constituent quark model approach with an
underlying $SU(2N_f) \otimes O(3)$ symmetry. This results in a number of new
predictions, which are consistent with but go beyond the Heavy Quark
Symmetry predictions for the
reduced form factors that describe current-induced bottom
baryon to charm baryon transitions.

\section{Classification of $s$- and $p$-Wave Heavy Baryon States}
A heavy baryon is made up of a light diquark system $(qq)$ and a heavy
quark Q. The light diquark system has bosonic quantum numbers $j^P$ with
the total angular momentum $j=0,1,2 \dots $ and parity $P=\pm 1$. To each
diquark system with spin-parity $j^P$ there is a degenerate heavy baryon
doublet with $J^P=(j\pm1/2)^P$ ($j=0$ is an exception). It is important to
realize that the Heavy Quark Symmetry structure of the heavy baryon states
is entirely determined by the spin-parity $j^P$ of the light diquark
system.

{}From our experience with light baryons and light mesons we know that one
can get a reasonable description of the light particle spectrum in the
constituent quark model picture. This is particularly true for the
enumeration of states, their spins and their parities. As much as we
know up to now, gluon degrees of freedom do not seem to contribute to the
particle spectrum. It is thus quite natural to try the same constituent
approach to enumerate the light diquark states, their spins and their
parities. From the spin degrees of freedom of the two light
quarks one obtains a spin 0 and a spin 1 state. The total orbital state of
the diquark system is characterized by two angular degrees of freedom
which
we take to be the two independent relative momenta $k=\frac{1}{2}(p_{1}-
p_{2})$ and $K=\frac{1}{2}(p_{1}+p_{2}-2p_{3})$ which can be formed from the
two light quark momenta $p_{1}$ and $p_{2}$ and the heavy quark momentum
$p_{3}$ \cite{htk}. The k-orbital momentum describes relative orbital
excitations of
the two quarks, and the K-orbital momentum describes orbital excitations of
the center of mass of the two light quarks relative to the heavy quark.
The details are in the next section.
The $(k,K)$ basis is quite convenient for two reasons. First, Copley, Isgur
and Karl\cite{C6} have found that the $(k,K)$ basis diagonalizes the
Hamiltonian, when harmonic interquark forces are used. Second, it allows one
to classify the diquark states in terms of $SU(2N_f) \otimes O(3)$
representations.

Let us do just this for the two flavour case $N_f=2$ for both $s$- and
$p$-wave states.
\begin{itemize}
\item[a)] $s$-wave ground state $\sim
 \Yzehn_{\,10} \otimes \Yeins_{\,1}$\\
 The spin-flavour content of the $SU(4)$ diquark representation
 $\underline{10}$ can be determined by looking at how the $\underline{10}$
 representation decomposes under the $SU(2)_{spin} \otimes SU(2)_{flavour}$
 subgroup. One has
 \begin{equation}
 \Yzehn_{\,10}=\left( \;\; \Yanti_{\,1} \otimes \Yanti_{\,1} + \Yzehn_{\,3}
 \otimes \Yzehn_{\,3} \;\right)
 \end{equation}
 When coupling in the heavy quark one finally has the particle content
 \zeilea
 \bea
 \left[ q_1 q_2 \right]: \qquad 0^+ &\rightarrow&{\;\,\textstyle \frac{1}{2}^+}
  \quad \quad \! \Lambda_Q  \nonumber\\
 \left\{ q_1 q_2 \right\}:\qquad 1^+ &\nearrow \atop \searrow&
 \begin{array}{l}
 \frac{1}{2}^+\\
 \frac{3}{2}^+\\
 \end{array}\quad
 \begin{array}{l}
 \Sigma_Q\\
 \Sigma_Q^*\\
 \end{array}
 \eea
 \zeilee
\item[b)]$p$-wave ($l_k=0$, $l_K=1$) $\sim \Yzehn_{\,10} \otimes \YK_{\,3}$:
 The $\underline{10}$  decomposes as   before. The
 spin~0 and spin~1 pieces of the $\underline{10}$ couple with $l_K=1$ to
 give $j^P=1^-$ and $j^P=0^-,1^-,2^-$, respectively. One has the particle
 content

 \bea
 \left[ q_1 q_2 \right]:\qquad 1^- &\nearrow \atop \searrow&
   \left.\begin{array}{l}
   \frac{1}{2}^-\\
   \frac{3}{2}^-\\
   \end{array}
   \right\} \quad \{ \Lambda_{QK1}^{**} \}\nonumber\\
 \left\{ q_1 q_2 \right\}:\qquad 0^-&\rightarrow& \;\;
{\textstyle{\frac{1}{2}^-}}
   \quad\quad\;\;\Sigma_{QK0}^{**} \nonumber\\
 1^-&\nearrow \atop \searrow&
   \left.\begin{array}{l}
   \frac{1}{2}^-\\
   \frac{3}{2}^-\\
   \end{array}\right\}
   \quad \{ \Sigma_{QK1}^{**} \}\nonumber\\
 2^-&\nearrow \atop \searrow&
   \left.\begin{array}{l}
   \frac{3}{2}^-\\
   \frac{5}{2}^-\\   \end{array}\right\}
   \quad \{\Sigma_{QK2}^{**} \}\;.
 \eea
 \zeilee
\item[c)]$p$-wave ($l_k=1$, $l_K=0$) $\sim \Yanti_{\,6} \otimes \Yk_{\,3}$:
 The diquark  $\underline 6$ of $SU(4)$ decomposes under the
 $SU(2)_{spin} \otimes SU(2)_{flavour}$ subgroup as
 \begin{equation}
 \Yanti_{\,6}=\left(\;\; \Yanti_{\,1} \otimes \Yzehn_{\,3} + \Yzehn_{\,3}
 \otimes \Yanti_{\,1}\; \right)\;.
 \end{equation}
 After coupling with the $k$-orbital angular momentum $l_k=1$ and the heavy
 quark spin $s=1/2$, the particle content can then be determined to be

 \zeilea
 \bea
 \left[ q_1 q_2 \right]:\qquad 0^-&\rightarrow& \;\; {\textstyle{\frac{1}
   {2}^-}} \quad \quad \,\; \Lambda_{Qk0}^{**}\nonumber\\
 1^-&\nearrow \atop \searrow&
   \left.\begin{array}{l}
   \frac{1}{2}^-\\
   \frac{3}{2}^-\\
   \end{array}\right\}\quad
   \{ \Lambda_{Qk1}^{**}\} \nonumber\\
 2^-&\nearrow \atop \searrow&
   \left.\begin{array}{l}
   \frac{3}{2}^-\\
   \frac{5}{2}^-\\
   \end{array}\right\}\quad
   \{ \Lambda_{Qk2}^{**} \} \nonumber\\
 \left\{ q_1 q_2 \right\}:\qquad1^- &\nearrow \atop \searrow&
   \left.\begin{array}{l}
   \frac{1}{2}^-\\
   \frac{3}{2}^-\\
   \end{array}\right\}
   \quad \{ \Sigma_{Qk1}^{**} \}
 \eea
 \zeilee
\end{itemize}
One thus has altogether seven $\Lambda$-type p-wave states and seven
$\Sigma$-type $p$-wave states. The analysis can easily be extended to the
case $SU(6)\otimes 0(3)$ bringing in the strangeness quark in addition.

Let us mention that, in the charm sector, the states $\Lambda_c(2285)$ and
$\Sigma_c(2453)$ are well established while there is first evidence for
the $\Sigma_c^*(2510)$ state. Recently two excited states $\Lambda_c^{**}
(2593)$ and $\Lambda_c^{**}(2625)$ have been seen\cite{lambdac} which very
likely
correspond to the two $p$-wave states making up the $\{\Lambda_{cK1}^{**}\}$
Heavy Quark Symmetry doublet. The charm-strangeness states $\Xi_c(2470)$
and $\Omega_c(2720)$ have been seen and first evidence was presented for
the $\frac{1}{2}^+ \; \Xi'_c(2570)$ state with the flavour configuration
$c\{sq\}$. In the bottom sector, the $\Lambda_b(5640)$ has made its way
into the Particle Data Booklet listing while some indirect evidence has
been presented for the $\Xi_b(5800)$\cite{sigmab}.

\section{Heavy Baryon Wave Functions}
We start by defining the Bethe-Salpeter amplitude (wave function) of a
heavy baryon as
\beq
B_{\alpha \beta \gamma}(x_{1},x_{2},x_{3}) =\langle 0\vert T
\psi_{q_{1}\alpha}(x_{1})\psi_{q_{2}\beta}(x_{2})\psi_{Q\gamma}
(x_{3})\vert B,P\rangle\,,\label{bs}
\eeq
where $\vert B,P\rangle$ represents a particular heavy baryon state with
mass $M$ and momentum $P$. $\psi_{Q}$ represents the heavy quark field,
while the $\psi_{q}$'s represent the light quark fields. $\alpha, \beta$
and $ \gamma $ are Dirac indices.

In a recent paper \cite{ht} we showed, using the LSZ reduction theorem
and interpolating fields, along with the HQET, that the momentum space
Bethe-Salpeter
amplitude for a heavy baryon (i.e. the Fourier transform of (\ref{bs})),
can be written as
\beq
B_{\alpha \beta \gamma}(p_{1},p_{2},p_{3})=\chi_{\rho\delta\gamma}
(p_{1},p_{2},p_{3}) A^{\rho \delta}_{\alpha \beta}(p_{1},p_{2},p_{3})\,,
\label{bsf}
\eeq
where the $\chi$'s are projection operators satisfying the
Bargmann-Wigner equations
\beq
(\vslash -1)_{\gamma}\,^{\gamma'}\chi_{\rho \delta \gamma'}=0
\label{bw1}
\eeq
on each label. Here $v^{\mu}$ is the four velocity of the heavy baryon.
These project out particular spin and parity states from
the orbital wave functions $A$. We showed in [7] how to construct
these projection operators for arbitrary meson and baryon orbital resonances.
These $\chi$'s are reduced in terms of representations of ${\cal L}\otimes
O(3,1)$, where ${\cal L}$ and $O(3,1)$ , for baryons, are generated by
\beq
S_{\mu\nu}=\frac{1}{2}\sigma_{\mu\nu}\otimes \eins\otimes\eins+
\eins\otimes\frac{1}{2}\sigma_{\mu\nu}\otimes\eins+
\eins\otimes\eins\otimes \frac{1}{2}\sigma_{\mu\nu}
\eeq
and
\beq
L_{\mu\nu}=L_{\mu\nu}^{(k)}+L_{\mu\nu}^{(K)}\,,
\eeq
respectively where
\beq
L_{\mu\nu}^{(k)}=i(k_{\mu}\frac{\partial}{\partial k^{\nu}}-k_{\nu}
\frac{\partial}{\partial k^{\mu}})\,
\eeq
and similarly for $L_{\mu\nu}^{K}$. $S_{\mu\nu}$ acts on the Dirac
labels.
$L_{\mu\nu}^{(k)}$ is the angular momentum operator for the
relative orbital angular momentum of the light quark pair, while
$L_{\mu\nu}^{(K)}$ is the angular momentum operator for the relative
orbital angular momentum between the centre of mass (c.m.) of the light
quark pair and the heavy quark. Going to the rest frame of the baryon we
reduce $O(3,1)$ to the rotation group  $O(3)$. As we
see there are two contributions to the orbital angular momentum. To find
particular orbital angular momenta we look for appropriate eigenstates of
$\vec{L}^{2}$. On the other hand, the group $\cal L$ is reduced to
$SU(2)$ by use
of the Bargmann-Wigner equations on each of the Dirac labels and by specifying
the symmetry of these indices. Thus we end up with irreducible representations
of $SU(2)_{spin}\otimes O(3)_{orbital}$. In the basis that we have
chosen for the angular momenta, our procedure amounts to  first
adding the spin and orbital angular momenta of the light quarks to
obtain the light diquark spin $j$ and then adding in the Heavy Quark
spin to obtain a pair of degenerate baryons with $J=j+1/2$ and
$J=j-1/2$. The case $j=0$ is special. In this case there is only one heavy
baryon state with $J=1/2$, as, for example, for the $\Lambda$-type
baryon ground state $\Lambda_{Q}$.

The s-wave projection operators are now disposed of quite easily.
The $\chi_{\alpha\beta\gamma}$ for s-waves can only be functions of
$k_{\perp}^2$ or $K_{\perp}^2$. There are only two possibilites, either
antisymmetric or symmetric in the $\alpha\beta$ indices,
\beq
\chi_{[\alpha\beta]\gamma}^{\Lambda}=\chi_{\alpha\beta}^{0}(v)\Psi_{\gamma}
\label{bslambda}
\eeq
or
\beq
\chi_{\{\alpha\beta\}\gamma}^{\Sigma}=
\chi_{\alpha\beta}^{1,\mu}(v)\Psi_{\mu,\gamma}\,,
\label{bsigma}
\eeq
where $\chi^{0}_{\alpha\beta}(v)=
\frac{1}{2\sqrt{2}}[(\vslash +1)\gamma_{5}C]_{\alpha\beta}$
and
$\chi^{1,\mu}_{\alpha\beta}(v)=
\frac{1}{2\sqrt{2}}[(\vslash +1)\gamma^{\mu}_{\perp}C]_{\alpha\beta}$ with
$\gamma_{\perp}^{\mu}~=~\gamma^{\mu}~-~\vslash v^{\mu}$. Here $C$ is the
charge conjugation matrix. The normalisation is such that
$(\bar{\chi}^{0})^{\alpha\beta}\chi^{0}_{\alpha\beta}=1$ and
$(\bar{\chi}^{1,\mu})^{\alpha\beta}\chi^{1,\nu}_{\alpha\beta}=
-g^{\mu\nu}_{\perp}$, where $g^{\mu\nu}_{\perp}=g^{\mu\nu}-v^{\mu}v^{\nu}$.
The
Bargmann-Wigner equations (\ref{bw1}) ensure that \beq
v^{\mu}\Psi_{\mu}=0
\label{trans}
\eeq
and
\beq
(\vslash -1)\Psi=(\vslash -1)\Psi_{\mu}=0\,.
\label{bw2}
\eeq
The superscripts $\Lambda$ and $\Sigma$, in eqs. (\ref{bslambda}) and
(\ref{bsigma}), indicate that these correspond to
the projection operators of the $\Lambda$-type and $\Sigma$-type baryons
respectively. The meaning of $\chi^{0}_{\alpha\beta}$ and
$\chi^{1,\mu}_{\alpha\beta}$ is obvious. $\chi^{0}$ represents the
antisymmetric spin zero state of the light diquark, whereas $\chi^{1,\mu}$
represents the symmetric spin one state of the light diquark. The
$\Lambda$-type baryons are also antisymmetric in the light flavours ensuring
overall symmetry in $flavour\otimes spin\otimes space$  for the light
diquark. Similarly the $\Sigma$-type baryons are symmetric in the light
flavours. Then using the Bargmann-Wigner equations (\ref{bw2}) we find
that for the $\Lambda$ baryon the spinor $\Psi$ is the usual Dirac
spinor i.e.
\beq
\Psi_{\gamma}=u_{\gamma}
\eeq
whereas for the $\Sigma$-type spinor we see that (\ref{bw2}) is solved by
\begin{equation}
\Psi_{\mu,\gamma}=
\left\{{u_{\mu,\gamma}}\atop
{\frac{1}{\sqrt{3}}(\gamma_{\perp\mu}\gamma_{5}u)_{\gamma}}\right\}\,.
\end{equation}
This corresponds to the Heavy Quark Symmetry $\Sigma$ baryon doublet,
${\frac{1}{2}}^{+}$ and ${\frac{3}{2}}^{+}$. $u_{\mu,\gamma}$ is the
Rarita-Schwinger spinor.

Going back to eq. (\ref{bsf}) we can now write the $\Lambda_{Q}$ s-wave
baryon wave function as
\beq
B_{\alpha \beta \gamma}=\phi_{\alpha \beta}(v,k,K)u_{\gamma}\,,
\eeq
where $\phi_{\alpha
\beta}(v,k,K)=\chi^{0}_{\rho\delta}(v)A_{\alpha\beta}^{\rho\delta}(v,k,K) $.

In the same manner the wave
function for the $\Sigma$-type s-wave heavy quark symmetry doublet can
be written as
\begin{equation}
B_{\alpha\beta\gamma}=\phi^{\mu}_{\alpha\beta}(v,k,K)
\left\{{u_{\mu,\gamma}}\atop
\frac{1}{\sqrt{3}}(\gamma_{\perp\mu}\gamma_{5}u)_{\gamma}\right\}
\end{equation}
where
\beq
\phi^{\mu}_{\alpha\beta}(v,k,K)=
\chi^{1,\mu}_{\delta\rho}(v)A^{\delta\rho}_{\alpha\beta}(v,k,K)\,.
\eeq

For p-waves the eigenvalue equation to be solved is
\beq
\vec{L}^{2}\chi=2\chi\,.
\eeq
There are two possibilities here. Either $(l_{k}=1,l_{K}=0)$ or
$(l_{k}=0,l_{K}=1)$.
Thus the above equation
is solved either by
\beq
\chi(k)=k_{\perp\lambda}\bar{\chi}^{\lambda}\,.
\eeq
or
\beq
\chi(K)=K_{\perp\lambda}\bar{\chi}^{\lambda}\,.
\eeq
Here
\beq
k^{\lambda}_{\perp}=k^{\lambda}-v\cdot k v^{\lambda}
\eeq
such that
\beq
v\cdot k_{\perp}=0\,.
\eeq
 Thus restituting the Dirac labels, the p-wave baryon projection
operators are of the form
$\chi_{\alpha\beta\gamma}=k_{\perp\lambda}\bar{\chi}^{\lambda}_{\alpha \beta
\gamma}$
or $\chi_{\alpha\beta\gamma}=K_{\perp\lambda}\bar{\chi}^{\lambda}_{\alpha
\beta
\gamma}$. Note also that although both $k$ and $K$ are of mixed symmetry
(under the interchange of $p_{1},p_{2},p_{3}$), k is symmetric under
$p_{1}\leftrightarrow p_{2}$ whereas $K$ is symmetric. We then construct
$\bar{\chi}^{\lambda}_{\alpha\beta\gamma}$ using the basic building blocks
$\chi^{0}_{\alpha\beta}$ and $\chi^{1,\lambda}_{\alpha\beta}$, alongwith
Dirac and generalised Rarita-Schwinger spinors, and ensuring
overall symmetry with respect to $flavour\otimes spin \otimes orbital$
for the diquarks.

A similar procedure can be followed for any orbital resonance where
the eigenvalue equation to be solved is $\vec{L}^{2}\chi=l(l+1)\chi$.
This has been worked out in detail in Ref. [7]. See also Ref. [8].
We now write down, in general, the covariant spin wave functions
$B_{\alpha\beta\gamma}$ of the Heavy Quark Symmetry baryon doublets
with $J=j+1/2$ and $J=j-1/2$
corresponding to the diquark spin $j$.
One has \cite{htk,C7}

\begin{equation}
B_{\alpha\beta\gamma}=\phi^{\mu_1\cdots\mu_j}_{\alpha\beta}(v,k,K)
\left\{N_{j}(\gamma_{\perp\mu_1}\gamma_{5}u_{\mu_2\cdots\mu_j})_{\gamma}\atop
N^{'}_{j}u_{\mu_1\cdots\mu_j;\gamma}\right\}
\label{eqa}
\end{equation}
where we have explicitly written out the Dirac spinor indices $\alpha$,
$\beta$ and $\gamma$. As before the spinor indices $\alpha$ and $\beta$
refer to
the light quark system and the index $\gamma$ refers to the heavy quark.
Here
\beq
\phi^{\mu_1\cdots\mu_j}_{\alpha\beta}(v,k,K)=
{\hat \phi}_{\delta\rho}^{\mu_1\cdots\mu_j}(v,k,K)
A^{\delta\rho}_{\alpha\beta}(v,k,K)\,,
\eeq
where ${\hat \phi}$ are listed in Table~\ref{tab1} for $p$-waves. The
general ${\hat \phi}$'s can be found in Ref.[7]. For the $s$-wave heavy
$\Lambda$ it is clear that ${\hat
\phi}_{\alpha\beta}=\chi^{0}_{\alpha\beta}$ and similarly for the
$s$-wave heavy $\Sigma$ we have  ${\hat
\phi}^{\mu}_{\alpha\beta}=\chi^{1,\mu}_{\alpha\beta}$.

Dropping the spinor indices one has

\beq
B=\phi^{\mu_1\cdots\mu_j}\Psi_{\mu_1\cdots\mu_j}
\label{eqb}
\eeq
where the ``superfield'' heavy baryon wave function $\Psi_{\mu_1\cdots\mu_j}$
stands for the two spin wave functions $\{j-1/2,j+1/2\}$ as indicated in
Eq.(\ref{eqa}).  The
normalizations ($N_{j}$ and $N'_{j}$)are fixed by the normalization condition
\beq
\bar\Psi^{\mu_1\cdots\mu_j}\Psi_{\mu_1\cdots\mu_j}=(-1)^{J-1/2}2M
\eeq
which gives $N_0=0$, $N_1=\sqrt{1/3}$, $N_2=\sqrt{1/10}$ and $N'_0=0$,
$N'_1=N'_2=1$ for the $s$- and $p$-wave cases discussed in this paper.
There is an implicit understanding that the set of tensor indices
``$\mu_1\cdots\mu_j$'' is always completely symmetrized, traceless with
regard to any pair of indices and transverse to the line of flight (i.e.
to $v^{\mu}$) in every
index. This is shown explicitly in Table~\ref{tab1} where
the $s$- and $p$-wave heavy baryon wave functions are listed. For example,
the notation  $\{\mu_1^\perp\mu_2^\perp\}_0$ implies symmetrization,
tracelessness and transversity of the two tensor indices $\mu_1\mu_2$ as
specified above. This specification is already implicit in the definition of
the Rarita-Schwinger spinors $u_{\mu_1 \dots \mu_j}$ and is therefore not
written out explicitly in these cases.
\begin{table}
\tcaption{\label{tab1}Spin wave functions (s.w.f.) of heavy $\Lambda$-type
      and $\Sigma$-type $s$-
      and $p$-wave heavy baryons.}
\vspace{5mm}
\renewcommand{\baselinestretch}{1.2}
\small \normalsize
\begin{center}
\begin{tabular}{|c|cccc|}
\hline \hline
& \begin{tabular}{c}
    light side s.w.f.\\
    ${\hat\phi}^{\mu_{1} \dots \mu_{j}}$
  \end{tabular}
& $j^{P}$
& \begin{tabular}{c}
    heavy side s.w.f.\\
    $\Psi_{\mu_{1} \dots \mu_{j}}$
  \end{tabular}
&$J^{P}$
\\ \hline \hline
\multicolumn{5}{|l|}{$s$-wave states ($l_k=0,l_K=0$)}\\
$\Lambda_{Q}$&$\hat \chi$&$ 0^{+}$&$u$&$\frac{1}{2}^{+}$\\
\hline
$\{ \Sigma_{Q} \}$ &${\hat \chi}^{1\mu_{1}}$& $1^{+}$&
	$\begin{array}{r}
	     \frac{1}{\sqrt{3}}\gamma^{\perp}_{\mu_{1}} \gamma_{5}u\\
	     u_{\mu_{1}}
	 \end{array} $
       &$\begin{array}{c}
	     \frac{1}{2}^{+}\\
	     \frac{3}{2}^{+}
	 \end{array} $
\\ \hline\hline
\multicolumn{5}{|l|}{$p$-wave states ($l_k=0,l_K=1$)}\\
$\{ \Lambda_{QK1}^{**} \} $&$ {\hat \chi}^{0} K_{\perp}^{\mu_{1}}$&
$1^{-}$&
    $\begin{array}{r}
      \frac{1}{\sqrt{3}} \gamma^{\perp}_{\mu_{1}} \gamma_{5} u\\
      u_{\mu_{1}}
     \end{array} $
   &$\begin{array}{c}
     \frac{1}{2}^{-} \\
     \frac{3}{2}^{-}
     \end{array}$
\\ \hline
$\Sigma_{QK0}^{**} $&$\frac{1}{\sqrt{3}}{\hat\chi}^{1} \cdot
K_{\perp}$&$0^{-}$&$u$&$
\frac{1}{2}^{-}$
\\ \hline
$\{ \Sigma_{QK1}^{**} \} $&$ \frac{i}{\sqrt 2} \varepsilon(\mu_{1} {\hat \chi}
^{1} K_{\perp}v)$&$1^{-}$&
       $\begin{array}{r}
	    \frac{1}{\sqrt{3}} \gamma^{\perp}_{\mu_{1}} \gamma_{5} u\\
	    u_{\mu_{1}}
       \end{array}$
     &$\begin{array}{c}
	    \frac{1}{2}^{-} \\
	    \frac{3}{2}^{-}
       \end{array}$
\\ \hline
$ \{ \Sigma_{QK2}^{**} \}$&$ \frac{1}{2} \{ {\hat \chi}^{1,\mu_{1}} K_{
\perp}^{\mu_{2}} \}_{0}$&$2^{-}$&
     $\begin{array}{r}
	   \frac{1}{\sqrt{10}}\gamma_{5} \gamma^{\perp}_{ \{ \mu_{1} }
				     u_{\mu_{2} \}_{0} }^{\hspace{1mm}}\\

	   u_{\mu_{1} \mu_{2}}
      \end{array}$
    &$\begin{array}{c}
	  \frac{3}{2}^{-} \\
	  \frac{5}{2}^{-}
      \end{array}$
\\ \hline \hline
\multicolumn{5}{|l|}{$p$-wave states ($l_k=1,l_K=0$)}\\
$\{ \Sigma_{Qk1}^{**} \}$&${\hat \chi}^{0} k^{\mu_{1}}_{\perp}$&$1^{-}$&
       $\begin{array}{r}
	    \frac{1}{\sqrt{3}}\gamma^{\perp}_{\mu_{1}} \gamma_{5} u\\
	    u_{\mu_{1}}
	\end{array}$
     &$\begin{array}{c}
	    \frac{1}{2}^{-}\\
	    \frac{3}{2}^{-}
	\end{array}$
\\ \hline
$\Lambda_{Qk0}^{**}$&$\frac{1}{\sqrt{3}}{\hat \chi}^{1} \cdot
k_{\perp}$&$0^{-}$&$u$&$
\frac{1}{2}^{-}$
\\ \hline
$\{ \Lambda_{Qk1}^{**} \}$&$\frac{i}{\sqrt 2} \varepsilon (\mu_{1} {\hat
\chi}^{1} k_{\perp} v)$&$1^{-}$&
    $ \begin{array}{r}
      \frac{1}{\sqrt{3}} \gamma^{\perp}_{\mu_{1}}\gamma_{5}u\\
      u_{\mu_{1}}
      \end{array} $
  &$  \begin{array}{c}
      \frac{1}{2}^{-} \\
      \frac{3}{2}^{-}
      \end{array} $
\\ \hline
$ \{ \Lambda_{Qk2}^{**} \} $&$ \frac{1}{2} \{ {\hat \chi}^{1,\mu_{1}}
k_{\perp}^{\mu_{2}} \}_{0}$&$2^{-}$&
    $\begin{array}{r}
     \frac{1}{\sqrt{10}} \gamma_{5} \gamma_{\{ \mu_{1}}^{\perp}
			       u_{\mu_{2} \}_{0} }^{\hspace{1mm}}\\
     u_{\mu_{1} \mu_{2} }
     \end{array} $
  &$\begin{array}{c}
    \frac{3}{2}^{-}\\
    \frac{5}{2}^{-}
    \end{array}$
\\ \hline \hline
\end{tabular}
\renewcommand{\baselinestretch}{1}
\small \normalsize
\end{center}
\end{table}

The meaning of $\phi^{\mu_{1}\cdots\mu_{j}}$ is clear. It stands for the
spin wave function of the light diquark with spin $j$. One can loosely
call the $\Psi_{\mu_1\cdots\mu_j}$ the heavy side spin wave function.
As mentioned before, $\phi^{\mu_{1}\cdots\mu_{j}}$ need not be
specified any further, i.e. we do not need to know
$A_{\alpha\beta}^{\delta\rho}(v,k,K)$, if we only wish to derive the
consequences of the
Heavy Quark Symmetry. In general these functions
$A_{\alpha\beta}^{\delta\rho}(v,k,K)$ are different for
different values of $j$. However, in the constituent quark model
with
$SU(2N_{f})\otimes O(3)$ symmetry, we have
$\phi^{\mu_1 \dots \mu_j} \rightarrow {\hat \phi}^{\mu_1 \dots \mu_j}$
\cite{C7} i.e.
$A_{\alpha\beta}^{\delta\rho}(v,k,K)=
\delta_{\alpha}^{\delta}\delta_{\beta}^{\rho}$.

{}From Table~\ref{tab1} we see that, for the $p$-wave spin wave
functions, we can further
explicitly pull out the relative momentum factor to write
\beq
{\hat \phi}^{\mu_{1}\dots\mu_{j}}(v,p)= {\hat
\phi}^{\mu_{1}\dots\mu_{j};\lambda}(v)p^{\perp}_{\lambda}\,,
\eeq
where $p^{\perp}_{\lambda}=k^{\perp}_{\lambda}$ or $K^{\perp}_{\lambda}$.
Now
${\hat \phi}^{\mu_{1}\dots\mu_{j};\lambda}(v)$ is only a function of $v$.
Although we are only dealing with diquark spins $j=0,1,2$ in this paper
the generic notation turns out to be quite convenient even for these
simple cases. For the $p$-wave light-side spin wave functions one has the
normalisation conditions
\beq
\left(\bar{\hat{\phi}}^{\nu_{1}\dots\nu_{j};\lambda}\right)^{\alpha\beta}
\left(\hat{\phi}^{\mu_{1}\dots\mu_{j};\lambda^{'}}\right)_{\alpha\beta}
g_{\lambda\lambda^{'}}^{\perp}=G^{\nu_{1}\dots\nu_{j};\mu_{1}\dots\mu_{j}}
\label{norm}
\eeq
where $G^{\nu_{1}\dots\nu_{j};\mu_{1}\dots\mu_{j}}$ is a generalised
transverse metric tensor\cite{htk}. Fro example, for $j=0,1,2$ one has
\bea
j=0 && G=1\\
j=1 && G^{\mu\nu}=-g^{\mu\nu}_{\perp}\\
j=2 && G^{\nu_{1}\nu_{2};\mu_{1}\mu_{2}}=
\frac{1}{2}(g_{\perp}^{\nu_{1}\mu_{1}}g_{\perp}^{\nu_{2}\mu_{2}}+
g_{\perp}^{\nu_{1}\mu_{2}}g_{\perp}^{\nu_{2}\mu_{1}}
-\frac{2}{3}g_{\perp}^{\nu_{1}\nu_{2}}g_{\perp}^{\mu_{1}\mu_{2}})
\eea
One sees that the generic form of the normalisation condition
(\ref{norm}) also covers the case of the $s$-wave normalisation with a
slight adaption of notation.

For the full $SU(2N_f)$ wave functions one also needs to avail of the
flavour diquark wave functions. For example, in the two flavour case $(u,d)$
one has the flavour wave functions
\beqn
D(I=0; I_3=0)=\frac{1}{\sqrt{2}} (ud-du), \qquad D(I=1;I_3=+1)=uu,
\eeqn
\beqn
D(I=1; I_3=0)=\frac{1}{\sqrt{2}} (ud+du), \qquad D(I=1;I_3=-1)=dd.
\eeqn
For the applications we have in mind the corresponding flavour-space
contractions are so simple that they are not listed separately. Also the
extension to the three-flavour case $(u,d,s)$ is straightforward.

\section{Coupling Structure of Current Transitions}
 The heavy-side and light-side transitions for $b\rightarrow c$
transitions occur
completely independent of each other (they ``factorize'') except for
the requirement that the heavy side and the light side have the same
velocity in the initial and final state, respectively, which are also
the velocities of the initial and final heavy baryons. The
$b\rightarrow c$ current transition, induced by the flavour-spinor
matrix~$\Gamma$, is hard in general and accordingly there is a change
of velocities $v_1\rightarrow v_2$. The heavy-side transitions
are completely specified whereas the light-side transitions
$j_1^{p_1}\rightarrow j_2^{p_2}$ are
described by a number of form factors or coupling factors which
parametrize the light-side transitions.

Now it is easy to write down the generic expressions for the current
transitions following from the Heavy Quark symmetry limit \cite{htk,C7}.
One has

\begin{equation}\label{cur trans}
\bar\Psi^2_{\nu_1\cdots\nu_{j_2}}\Gamma\Psi_{\mu_1\cdots\mu_{j_1}}^1
\left(\sum_{i=1}^Nf^i(\omega)
t_i^{\nu_1\cdots\nu_{j_2};\mu_1\cdots\mu_{j_1}}\right)
\label{eqd}
\end{equation}
where the $\Psi_{\mu_1\cdots\mu_j}$ are the heavy baryon spin wave
functions introduced in Sec.~1. The tensors
$t_i^{\nu_1\cdots\nu_{j_2};\mu_1\cdots\mu_{j_1}}$ describe the light
side transitions. Here $\omega$ is the velocity transfer variable defined
as $\omega=v_1\cdot v_2$.

Thus the counting of the number of reduced form factors $f^i(\omega)$, that
describe the heavy baryon transitions,
can readily be done by referring to the number N of independant diquark
transition amplitudes $t_i^{\nu_1\cdots\nu_{j_2};\mu_1\cdots\mu_{j_1}}$
\cite{C7}.
Defining the normality n of a diquark state with
quantum numbers $j^{P}$ by $n=P(-1)^{j}$ one has to differentiate between
the two cases where the product of normalities of the two diquark states
is even or odd. One finds
\begin{eqnarray}
n_1\cdot n_2&=&1\qquad N=j_{\mini}+1\nonumber\\
n_1\cdot n_2&=&-1\quad N=j_{\mini}\nonumber
\end{eqnarray}

The generic expression Eq.(\ref{eqd}), completely determines the Heavy
Quark Symmetry structure of the current transitions.
What remains to be done is to write down independent sets of
covariant coupling tensors.  One notes that
the heavy baryon transition amplitudes are factorized into a known
heavy-side transition and an unknown light-side transition.  We might
add that a factorization of similar nature also occurs in other areas
of particle physics as e.g. in the case of lepton-hadron interactions
where one has a current-current interaction $j_{\mu}^{lepton}
J^{\mu}_{hadron}$, and where the lepton current is known and the
hadron current is parametrized by a set of unknown form factors.

The current transition tensors $t_i^{\nu_1\cdots\nu_{j_2};
\mu_1\cdots\mu_{j_1}} $ in Eq.(\ref{cur trans}) have to be built from
the vectors $v_1^{\nu_i}$ and $v_2^{\mu_i}$, the metric tensor
$g_{\mu_i\mu_k}$ and, depending on parity, from the Levi-Civita
object $\varepsilon(\mu_i\nu_kv_1v_2):=\varepsilon_{\mu_i\nu_k\alpha\beta}
v_1^\alpha v_2^\beta$. The relevant basic tensors are
listed in Table~\ref{schwache Uebergaenge}.

\begin{table}
\bec
\tcaption{\label{schwache Uebergaenge}Tensor structure of the diquark
transitions $j_1^{P_1} \rightarrow j_2^{P_2}$. Sign of the product of
normalities $n_1 \cdot n_2$ and the smaller value of $j_1, \, j_2$
determine the number N of independent transitions
or Isgur-Wise functions as specified in Eq.(\protect{\ref{cur trans}}) }
\vspace{5mm}
\renewcommand{\baselinestretch}{1.5}
\small \normalsize
\begin{tabular}{|lr|cc|l|c|l|}
\hline
\multicolumn{2}{|c|}{diquark transition} & & &  covariant amplitude \\
\hline
$j_1^{P_1} \rightarrow$ & $j_2^{P_2}$ & $n_1 \cdot n_2$ & $N$ &
  $t^{\nu_1 \dots \nu_{j_2};\mu_1 \dots \mu_{j_1}}$ \\
\hline\hline
$0^+ \rightarrow$ & $0^+$ & $+1$ & $1$ & $f^{(0)} \cdot 1$ \\
\hline
$0^+ \rightarrow$ & $0^-$ & $-1$ & $0$ & - \\
& $1^-$ & $+1$ & $1$ & $f_{1,2}^{(1)} \cdot v_{1\nu_1}$ \\
& $2^-$ & $-1$ & $0$ & -\\
\hline \hline
$1^+ \rightarrow$ & $1^+$ & $+1$ & $2$ & $-g_1^{(0)} \cdot g_{\mu_1 \nu_1}+
g_2^{(0)} \cdot v_{1\nu_1} v_{2\mu_1}$\\
\hline
$1^+ \rightarrow$ & $0^-$ & $+1$ & $1$ & $g_2^{(1)} \cdot v_{2\mu_1}$\\
& $1^-$ & $-1$ & $1$ & $ig_{1,3}^{(1)} \cdot \varepsilon(\mu_1\nu_1 v_1 v_2)$
\\
& $2^-$ & $+1$ & $2$ & $-g_4^{(1)} \cdot v_{1\nu_1} g_{\nu_2\mu_1}+
g_5^{(1)} \cdot v_{1\nu_1}v_{1\nu_2}v_{1\mu_1}$ \\
\hline \hline
\end{tabular}
\renewcommand{\baselinestretch}{1}
\small \normalsize
\eec
\end{table}
The number N of independent form factors for a given diquark
transition can easily be obtained with the help of Eq.(\ref{cur
trans}) and is also listed in Table \ref{schwache Uebergaenge}. The
column ``covariant amplitudes'' contains the basic tensors and the
invariant or reduced amplitudes that multiply them. The nomenclature
is such that the quasi-elastic $\Lambda_b \rightarrow \Lambda_c$
reduced form factor (or Isgur-Wise function) is denoted by
$f^{(0)}(\omega)$ with the zero recoil normalization condition
$f^{(0)}(1)=1$. The transitions to the $p$-wave states $\Lambda_b
\rightarrow \{\Lambda_{cK1}^{**}\}$ and $\Lambda_b
\rightarrow \{\Lambda_{ck1}^{**}\}$ are described by the form factors
$f_1^{(1)}(\omega)$ and $f_2^{(1)}(\omega)$, respectively. For the
$\Sigma$-type transitions there are two form factors each for the
$\Sigma_b \rightarrow \{\Sigma_c\}$ and $\Sigma_b \rightarrow
\{\Sigma_{cK2}^{**}\}$ transitions involving the diquark tarnsitions
$1^+ \rightarrow 1^+$ and $1^+ \rightarrow 2^-$, respectively. The
quasi-elastic $\Sigma_b \rightarrow \Sigma_c$ form factor
$g_1^{(0)}(\omega)$ is normalized to 1 at zero recoil. The form
factors $g_1^{(1)}(\omega)$ and $g_3^{(1)}(\omega)$, finally,
describe the $\Sigma_b \rightarrow \Sigma_{ck1}^{**}$ and $\Sigma_b
\rightarrow \Sigma_{cK1}^{**}$ transitions.

As an application we shall write down the contributions of the
baryonic $s$- and $p$-wave states to the sum rule of Bj\o rken\cite{C9}.
The calculation is straightforward yet cumbersome \cite{htk}. It involves the
use of the covariance structure listed in Table \ref{schwache Uebergaenge}
as well as the heavy baryon wave functions listed in Table \ref{tab1}. In
the case of the $\Lambda_b \rightarrow \Lambda_c,\; \Lambda_c^{**}$
transitions one finds
\beq\label{lambda Summe}
1=|f^{(0)}(\omega)|^2_{L=0} +(\omega^2-1) \left\{|f_1^{(1)}|^2_{L=1}
+|f_2^{(1)}(\omega)|^2_{L=1}\right\} + \dots
\eeq
where the ellipsis stand for the contributions of higher radial and
orbital excitations not considered here, and for continuum contributions.
For the $\Sigma$-type transitions $\Sigma_b \rightarrow \Sigma_c, \Sigma_c^*,
\Sigma_c^{**}$ one has
\bea
1&=& \frac{1}{9} | (\omega +2) g_1^{(0)}(\omega)-(\omega^2-1)
  g_2^{(0)} (\omega)|^2_{L=0}\nonumber\\
&&+\frac{2}{9} (\omega-1)^2 |g_1^{(0)}(\omega) -(\omega +1)
  g_2^{(0)}(\omega)|^2_{L=2} \nonumber\\
&&+(\omega^2-1) \{ \frac{2}{3} |g_1^{(1)}|^2_{L=1} +\frac{1}{3}
  |g_2^{(1)}|^2_{L=1} + \frac{4}{3} |g_3^{(1)}(\omega)|^2_{L=1} \nonumber\\
&&\qquad +\frac{8}{45}|(\omega +\frac{3}{2}) g_4^{(1)} -(\omega^2 -1)
  g_5^{(1)} |^2_{L=1}\nonumber\\
&&\qquad +\frac{12}{45} (\omega-1)^2 |g_4^{(1)}(\omega)-(\omega+1)
  g_5^{(1)}|^2_{L=3} \} \dots \label{Sigsum}
\eea
For the $\Sigma_b \rightarrow \{ \Sigma_c, \Sigma_c^* \} $ and the $\Sigma_b
\rightarrow \{ \Sigma_{cK2}^{**} \}$ transitions we have used a
diagonal basis in terms of the partial wave amplitudes as indicated in
Eq.(\ref{Sigsum}). The subscript $L$ in equations (\ref{lambda Summe}) and
(\ref{Sigsum}) refer to the different partial wave contributions.
We see that the contributions of the various $\Lambda$-type and
$\Sigma$-type states exhibit the characteristic
$p^{2L} \propto (\omega-1)^L$ threshold powers of the diquark transitions.

The presentation of the Bj\o rken sum rules completes our discussion of
the Heavy Quark Symmetry structure of current-induced heavy baryon
transitions.  The set of reduced form factors constitute the maximal
possible reduction of the full complexity of the form factor problem
which can be achieved using just the Heavy Quark Symmetry. To obtain further
information on the reduced form factors one necessarily has to study
the internal dynamics of the light diquark system itself. This is a
notoriously difficult problem since it involves the solution of
quark-gluon dynamics in the nonperturbative regime. First attempts at
solving this problem in the baryonic sector are being undertaken
using lattice techniques, QCD sum rule methods, or, more
conventionally, using some potential type quark model. Short of doing
a full-fledged dynamical calculation one can work in the context of
the constituent quark model and try to extract as much information on
the diquark transitions using the $SU(2N_f)\otimes O(3)$ symmetry
classification of the diquark states. This is the subject of the next
section.

\section{$SU(2N_f) \otimes O(3)$ Structure of the Light-Side
Transitions}
For the $s$-wave to $s$-wave transition
the relevant matrix element for the light side transition is given by
\beq
\sum_{i=1}^Nf^i(\omega)
t_i^{\nu_1\cdots\nu_{j_2};\mu_1\cdots\mu_{j_1}}=
\left(\bar{\hat \phi}^{\nu_1 \dots \nu_{j_2}}(v_{2})\right)^{\alpha' \beta'}
({\cal O} )_{\alpha' \beta'}^{\alpha \beta}\left({\hat \phi}^
{\mu_1 \dots \mu_{j_1}}(v_1)\right)_{\alpha \beta}\,,\label{s-s}
\eeq
where the ${\hat \phi}$'s are either $\chi^{0}$ or $\chi^{1,\mu}$, i.e.
they either have zero indices or just one index, depending on whether we
have a $\Lambda_{b}$ to $\Lambda_{c}$ transition  or $\Sigma_{b}$ to
$\Sigma_{c}$ transition. ${\cal O}$ is the operator given by the overlap
integral
\beq
{\cal O}_{\alpha' \beta'}^{\alpha \beta}
=\int d^{4}k_{2}d^{4}K_{2}d^{4}k_{1}d^{4}K_{1}
\bar{A}_{\alpha'\beta'}^{\rho\sigma}{\cal T}_{\rho\sigma}^{\rho'\sigma'}
A_{\rho'\sigma'}^{\alpha \beta}\,.\label{O}
\eeq
Here ${\cal T}_{\rho\sigma}^{\rho'\sigma'}(k_{2},K_{2};k_{1},K_{1})$ is
the transition kernel. $k_{1},K_{1}$ and $k_{2},K_{2}$ are repectively
the relative momenta $k,K$ for the incoming and outgoing quarks.
This overlap integral is common for both $\Lambda$ and $\Sigma$ type
transitions but this information is not of much help. It can be seen,
from general arguments of the Lorentz structure, that ${\cal O}$
in general consists of five independent contributions \cite{hkkt}
$\eins\otimes\eins$, $\gamma_{5}\otimes\gamma_{5}$,
$\gamma_{\mu}\otimes\gamma^{\mu}$,
$\gamma_{\mu}\gamma_{5}\otimes\gamma^{\mu}\gamma_{5}$ and
$\sigma_{\mu\nu}\otimes\sigma^{\mu\nu}$\footnote{Notice that there are
no $\vslash_1$ or $\vslash_2$ terms as these reduce to $\eins\;$ when acting on
the projectors.}, which we call the $S,P,V,A$ and $T$ couplings of the spins
of the two spectator systems of diquarks. Here the $S$ operator is a one-body
operator whereas the others are two-body operators involving
interactions between the light quarks. This set can be divided into the three
antisymmetric independent combinations $2S-\frac{1}{2}T+A$, $V+A$ and
$S-V-P$ contributing to the $\Lambda$ transitions and the two symmetric
combinations $2S+V-A-2P$ and $3S+\frac{1}{2}T+3P$ contributing to the
$\Sigma$ transitions \cite{hkkt}. In this way way we get no relation
between the $\Lambda$ and $\Sigma$ transition form factors.

Similarly for the transitions from the ground state bottom baryon to the
$p$-wave charm
baryon states the relevant light side matrix element is given by
\beq\label{s-p trans}
\left( \bar{\hat \phi}^{\nu_1 \dots \nu_{j_2};\nu}(v_2)\right)^{\rho' \sigma'}
({\cal O}_{\nu}^p)_{\rho' \sigma'}^{\rho \sigma}
\left({\hat \phi}^{\mu_1 \dots \mu_{j_1}}(v_1)\right)_{\rho \sigma}\,.
\eeq
Here ${\cal O}_{\nu}^p$ is either  ${\cal O}^{K}_{\nu}$ or
${\cal O}^{k}_{\nu}$ with
\beq
({\cal O}_{\nu}^p)_{\alpha' \beta'}^{\alpha \beta}
=\int d^{4}k_{2}d^{4}K_{2}d^{4}k_{1}d^{4}K_{1}\;p_{\nu}^{\perp}
\bar{A}_{\alpha'\beta'}^{\rho\sigma}{\cal T}_{\rho\sigma}^{\rho'\sigma'}
A_{\rho'\sigma'}^{\alpha \beta}\,,\label{OK}
\eeq
where $p_{\perp}$ is transverse to the outgoing velocity $v_2$. Here
again in general ${\cal O}_{\nu}^p$ can consist of the one-body
operators $v_{1\nu}^{\perp}\,
\eins\otimes\eins$ and
$(\eins\otimes\gamma^{\perp}_{\nu}\pm\gamma^{\perp}_{\nu}\otimes\eins\;)$ plus
a large number of two-body operators
$v_{1\nu}^{\perp}\gamma^{\mu}\otimes\gamma_{\mu}$, etc. This means that
we would not get any more relations beyond those already given in
Table~2.

However, recently interest in the constituent quark model has been rekindled
by the discovery (or rediscovery) that two-body spin-spin
interactions between quarks are nonleading in $1/N_C$, at least in
the baryon sector \cite{jmw,C10}. Thus, to leading order in $1/N_C$, light
quarks behave as if they were heavy and a classification of a light
quark system in terms of $SU(2N_f) \otimes O(3)$ symmetry multiplets
makes sense. In this case transitions between light quark systems are
parametrized only in terms of the set of one-body operators whose matrix
elements
are then
evaluated between the $SU(2N_f) \otimes O(3)$ multiplets.

For the diquark transitions discussed in this paper, the relevant
one-body operators are then given by

\noindent
$s$-wave to $s$-wave:

\beq\label{Op 1}
\mbox{\cal O}=A(\omega)\, \eins \otimes \eins
\eeq

\noindent
$s$-wave to $p$-wave: Here we distinguish transitions from the ground
state to the two types of p-wave excitations ($l_K=1$;$l_k=0$) and
($l_K=0$;$l_k=1$),  denoted by the superscripts $K$ and $k$ respectively,

\bea
{\cal O}^K_\nu=A^K(\omega) \, v_{1\nu}^{\perp}\, \eins\,
          \,\otimes \eins\,+ \, B_K(\omega)\,(\eins \otimes
          \gamma_{\nu}^{\perp}
+  \gamma_{\nu}^{\perp} \otimes \eins\,)\nonumber\\[4mm]
{\cal O}^k_\nu=A^k(\omega)\, v_{1\nu}^{\perp}\, \eins
          \,\otimes \,\eins\;
+ \, B_k(\omega)\,(\eins \otimes \gamma_{\nu}^{\perp} -
          \gamma_{\nu}^{\perp} \otimes \eins\,)\label{Op 2}
\eea

\noindent
 The reduced form
factors in Eqs.(\ref{Op 1}) and (\ref{Op 2}) depend on the velocity transfer
variable $\omega$ and are unknown functions expect for the
normalization condition $A(1)=1$ in Eq.(\ref{Op 1}).  The operators $\eins
\otimes \eins\,$ and $v_{1\nu}^{\perp} \;\eins \,\otimes \eins\,$ do not couple
angular and spin degrees of freedom. These were the operators used some thirty
years ago when the consequences of the collinear symmetry $SU(6)_W$ or
$\tilde U (12)$ were worked out. In the following we shall therefore refer
to the $A$-type operators as the collinear operators.  The operators $(\eins
\otimes \gamma^{\perp}_{\nu} \pm \gamma^{\perp}_{\nu} \otimes \eins\,)$ on the
other hand,
introduce spin-orbit coupling interactions and will therefore be called
spin-orbit operators.

The matrix elements, Eq.(\ref{s-s}) and Eq.(\ref{s-p trans}), of the  operators
Eq.(\ref{Op 1}) and
(\ref{Op 2}) can be readily evaluated using the light-side spin wave
functions in Table~\ref{tab1}.
The three ground state to ground state form factors $f^{(0)}(\omega),
\; g_1^{(0)}(\omega)$ and $g_2^{(0)}(\omega)$  can now
be expressed by the single form factor $A(\omega)$ as
\bea
f^{(0)}(\omega)  &=& \frac{\omega +1}{2} A(\omega)
                     \nonumber\\
g^{(0)}_1(\omega)&=& \frac{\omega +1}{2} A(\omega) \\
g^{(0)}_2(\omega)&=& \frac{1}{2} A(\omega)
                    \nonumber
\eea

This result has been derived before in Ref.[4] and agrees with the
large-$N_C$ predictions, reported in
Ref. [11], where a universal form factor $A(\omega)$ (with normalization
$A(1)=1$) determines the semileptonic
$\Lambda_b\rightarrow\Lambda_c$ and $\Sigma^{(*)}_b\rightarrow\Sigma^{(*)}_c$
transitions.
Looking at the Bj\o rken sum rule, Eq.(\ref{Sigsum}), it is clear, in
this case, that the
$\Sigma$-type transitions are predicted to result from pure $L=0$
diquark transitions. This is a testable prediction in as much as the
population of the helicity states in the decay baryon is fixed
resulting in a characteristic angular decay pattern of the subsequent
decays. More difficult is a test of the relation between the
$\Lambda$-type and the $\Sigma$-type form factors. In the test one
would have to compare $\Lambda_b \rightarrow \Lambda_c$ and $\Omega_b
\rightarrow \{\Omega_c, \; \Omega_c^*\}$ transitions (where there are
additional $SU(3)$ breaking effects) since these are the transitions that
are experimentally accessible. We mention that the $\Sigma_b \rightarrow
\{\Sigma_c, \; \Sigma_c^*\}$ branching fraction is expected to be too
small to be measurable.

For the $s$ to $p$ transitions the evaluation of the matrix element
(\ref{s-p trans}) is straightforward
and one obtains

\noindent
$(l_K=1; \; l_k=0)$
\bea
f_1^{(1)}(\omega) &=& \frac{\omega +1}{2} A^K(\omega) +
    B^K(\omega)\nonumber\\
g_2^{(1)}(\omega) &=& \frac{1}{\sqrt 3} \left( \frac{\omega +1}{2}
    A^K(\omega) + 3B^K(\omega)\right)\nonumber\\
g_3^{(1)}(\omega) &=& -\frac{1}{\sqrt 2} \left( \frac{\omega +1}{2}
    A^K(\omega) + B^K(\omega)\right)\label{K-multi}\\
g_4^{(1)}(\omega) &=& \frac{\omega +1}{2} A^K(\omega)\nonumber\\
g_5^{(1)}(\omega) &=& \frac{1}{2} A^K(\omega)\nonumber
\eea

\noindent
$(l_K=0;\; l_k=1)$
\bea
f_2^{(1)}(\omega)&=& \sqrt{2}  B^k(\omega) \nonumber\\
g_1^{(1)}(\omega)&=& - B^k(\omega)
\eea
These relations include but go beyond the Heavy Quark Symmetry relations
given in
Table~\ref{schwache Uebergaenge}. We see that the five form factors
$f_{1}^{(1)},g_{2-5}^{(1)}$ are
now expressed in terms of just two unknown functions whereas the two
form factors $f_{2}^{(1)}$ and $g_{1}^{(1)}$ are reduced to just one
unknown function $B^{k}$.

{}From the equation (\ref{Sigsum}) we see that
the ``partial wave" amplitudes for the case $1^+ \rightarrow 2^-$ are
related to the form factors through
\bea
A_{L=1}\propto (\omega
+\frac{3}{2})g_{4}^{(1)})-(\omega^{2}-1)g_{5}^{(1)}\nonumber\\
A_{L=3}\propto g_{4}^{(1)}-(\omega +1)g_{5}^{(1)}\,.
\eea
Putting in the expressions from Eqs.(\ref{K-multi}) we see that the
$A^{K}(\omega)$ contribution is purely $L=1$. One therefore predicts that
only the lower partial wave
$L=1$ is populated in the $\Sigma_b \rightarrow \{ \Sigma_{cK2}^{**}\}$
transition. As before this has implications for the population of
density matrix elements of the daughter baryons that can be
experimentally checked. In the collinear limit, where $B^k=0$ and $B^K =0$,
the transitions to the $p$-wave $k$-multiplet are completely
suppressed and there are simple relations for the transitions into
the $K$-muliplet. It would be interesting to experimentally determine
the relative strengths of the collinear and spin-orbit.
{}From naive expectations one would expect the spin-orbit contributions
to be suppressed relative to the leading collinear terms. Let us
finally remark that the transitions to the $k$- and $K$-multiplets
become related when one has a complete factorization of the light
quark wave functions (``independent quark motion'')\cite{hkkt}.

\section{Concluding Remarks}
We have studied the consequences of Heavy Quark Symmetry for current
transitions between heavy baryons. For these
transitions we discussed how the most general Heavy Quark Symmetry
structure can be further simplified by invoking a constituent quark
model $SU(2N_f) \otimes O(3)$ symmetry for the light side transition.
The key ingredient is that the transitions factorize
into a known heavy-side transition and an unknown light-side
transition when $m_Q \rightarrow \infty$. The
light-side and heavy-side transitions are linked through angular
coupling factors which are calculable. This factorization breaks down
when higher order $1/m_Q$ effects are taken into account. In this case
the light-side and heavy-side transitions become entangled in a very
nontrivial way. Depending on one's own attitude such additional
complications may be considered a curse or a blessing. From this
point of view the charm sector ($\Lambda_{QCD}/m_c \simeq (20-30)\%
$) is best suited for studying  {\cal O}($1/m_Q$) effects whereas
the bottom sector  ($\Lambda_{QCD}/m_b \simeq (6-10)\%$ ) would be
best suited to see how Heavy Quark Symmetry is at work.

\end{document}